\documentclass[submission,copyright,creativecommons]{eptcs}
\providecommand{\event}{ACL2 2023} 

\usepackage{iftex}
\usepackage{comment}
\usepackage{amsmath}
\usepackage{amsthm}
\usepackage{amssymb}
\usepackage[frozencache,cachedir=.]{minted}
\usepackage{xcolor} 
\definecolor{LightGray}{gray}{0.9}

\ifpdf
  \usepackage{underscore}         
  \usepackage[T1]{fontenc}        
\else
  \usepackage{breakurl}           
\fi

\newtheorem{theorem}{Theorem}[section]
\newtheorem*{theorem*}{Theorem}

\theoremstyle{definition}

\theoremstyle{remark}

\numberwithin{equation}{section}

\title{ACL2 Proofs of Nonlinear Inequalities with Imandra}
\author{Grant Passmore
\institute{Imandra Inc.\\ Austin, TX}
\institute{Clare Hall, Cambridge}
\email{grant@imandra.ai}
}
\def\titlerunning{ACL2 Proofs of Nonlinear Inequalities with Imandra}
\def\authorrunning{G.O.Passmore}
\begin{document}
\maketitle

\begin{abstract}
We present a proof-producing integration of ACL2 and Imandra for proving nonlinear inequalities. 
This leverages a new Imandra interface exposing its nonlinear decision procedures. 
The reasoning takes place over the reals, but the proofs produced are valid over the rationals and may be run in both ACL2
and ACL2(r).
The ACL2 proofs Imandra constructs are extracted from Positivstellensatz refutations, a real algebraic analogue of the Nullstellensatz, and are found using convex optimization.
%
\end{abstract}

\section{Introduction}

Nonlinear inequalities can pose critical formal verification challenges.
While nonlinear integer arithmetic is undecidable, nonlinear real arithmetic is decidable, and advances in decision procedures have brought many useful classes of problems within reach of automated methods. Unfortunately, most effective modern methods, e.g., those based on Cylindrical Algebraic Decomposition (CAD)~\cite{collins-cad}, are not \emph{proof producing} and rely on nontrivial computer algebra computations which must be trusted. This presents a major barrier for taking advantage of such techniques in formal proofs.

In this work, we use the Positivstellensatz~\cite{krivine-psatz,stengle-psatz}, a fundamental result in real algebraic geometry, to construct fully formal proofs of nonlinear real inequalities in ACL2.
The Positivstellensatz guarantees the existence of proofs of inequalities in a certain formal system, and advances in convex optimization (including semidefinite programming (SDP) and sums-of-squares decompositions) allow us to effectively search over a convex space of \emph{certificates} to find such proofs. When these proofs are found, we can then translate them into ACL2 proofs in a structured form that ACL2 can easily check.

Let us motivate our discussion with an example. Consider one direction of the discriminant criterion for solubility of a quadratic equation:
\[\forall x,a,b,c\in\mathbb{R}\left(ax^2 + bx + c = 0 \implies b^2 - 4ac \geq 0\right).\]
To prove this, we will negate and normalize its constraints s.t. all relations are drawn from $\{=,\geq,>,\neq\}$: 
\[ax^2 + bx + c = 0 \wedge 4ac - b^2 > 0\]
and then proceed to derive a contradiction. The Positivstellensatz (cf. Sec~\ref{sec:psatz}) guarantees the existence of a \emph{certificate} establishing unsatisfiability by a particularly simple form of argument. In this case, a certificate is given by
\[(4ac - b^2) + (2ax + b)^2 + (-4a)(ax^2 + bx + c)\]
as $(4ac - b^2)>0$ and $(-4a)(ax^2 + bx + c)=0$ by assumption, and $(2ax + b)^2\geq0$ as it is a square. Thus, by assumption, the certificate must be strictly positive. But by polynomial arithmetic, it is easy to verify that the certificate sums to $0$. Thus the negation of our conjecture implies that $0>0$. This is the general structure of a Positivstellensatz refutation, and all ACL2 proofs we produce proceed in this way.

\section{The Imandra-ACL2 Interface and ACL2 Proofs}

We have implemented the Positivstellensatz proof method in Imandra~\cite{passmore-imandra}, and built an interface which extracts ACL2 proofs from its refutations. 

To use it, one poses a conjecture as an S-expression in which all variables are implicitly taken to be reals. If Imandra is successful, an ACL2 proof is produced as an {\tt (ENCAPSULATE ...)} event which exports a single theorem named {\tt FINAL}. Note that in the ACL2 theorems we prove, the variables are {\tt RATIONAL} instead of {\tt REAL}, as {\tt REALP} is available only in ACL2(r). 

For our quadratic equation example, the input is:

\begin{minted}{lisp}
(IMPLIES (= (+ (* A X X) (* B X) C) 0)
         (>= (- (* B B) (* 4 A C)) 0))
\end{minted}

\noindent and the output (produced in a fraction of a second) is:

\begin{minted}{lisp}
(ENCAPSULATE ()

;; Preamble

 (SET-IGNORE-OK T)
 (SET-IRRELEVANT-FORMALS-OK T)

 (LOCAL (DEFMACRO NEQ (X Y)
          `(OR (< ,X ,Y) (> ,X ,Y))))

 (LOCAL (DEFUN SQUARE (X)
          (* X X)))

 (LOCAL (DEFTHM SQUARE-PSD
          (IMPLIES (RATIONALP X)
                   (>= (SQUARE X) 0))
          :RULE-CLASSES (:LINEAR)))

 (LOCAL (DEFTHM SQUARE-TYPE
          (IMPLIES (RATIONALP X)
                   (RATIONALP (SQUARE X)))
          :RULE-CLASSES (:TYPE-PRESCRIPTION)))

 (LOCAL (IN-THEORY (DISABLE SQUARE)))
 
 (LOCAL (include-book "arithmetic-5/top" :dir :system))

;; Normalized problem polynomials
 
 (LOCAL (DEFUND PROB-0 (A B C X)
          (+ (* A (* X X)) (+ (* B X) C))))
 
 (LOCAL (DEFUND PROB-1 (A B C X)
          (- 0 (- (* B B) (* 4 (* A C))))))

;; Normalized goal expressed using problem polynomials
 
 (LOCAL (DEFUN GOAL (A B C X)
          (IMPLIES (AND (RATIONALP A)
                        (RATIONALP B)
                        (RATIONALP C) (RATIONALP X))
                   (NOT (AND (= (PROB-0 A B C X) 0)
                             (> (PROB-1 A B C X) 0))))))

;; Ideal cofactors
 
 (LOCAL (DEFUND IDEAL-CF-0 (A B C X)
          (* -4 A)))
 
 (LOCAL (DEFTHM IDEAL-CF-0-TYPE
          (IMPLIES (AND (RATIONALP A)
                        (RATIONALP B)
                        (RATIONALP C) (RATIONALP X))
                   (RATIONALP (IDEAL-CF-0 A B C X)))
          :hints
          (("Goal" :in-theory (enable IDEAL-CF-0)))))

;; Cone cofactors
 
 (LOCAL (DEFUND CONE-CF-0 (A B C X)
          (SQUARE (+ (* 2 (* A X)) B))))
 
 (LOCAL (DEFTHM CONE-CF-0-TYPE
          (IMPLIES (AND (RATIONALP A)
                        (RATIONALP B)
                        (RATIONALP C) (RATIONALP X))
                   (RATIONALP (CONE-CF-0 A B C X)))
          :hints
          (("Goal" :in-theory (enable CONE-CF-0)))))
 
 (LOCAL (DEFTHM CONE-CF-0-PSD
          (IMPLIES (AND (NOT (GOAL A B C X))
                        (RATIONALP A)
                        (RATIONALP B)
                        (RATIONALP C) (RATIONALP X))
                   (>= (CONE-CF-0 A B C X) 0))
          :hints
          (("Goal" :in-theory
                   (enable CONE-CF-0 PROB-0 PROB-1)))
          :rule-classes (:linear)))

;; Monoid cofactors
 
 (LOCAL (DEFUND MONOID-CF-0 (A B C X)
          (- 0 (- (* B B) (* 4 (* A C))))))

;; Positivstellensatz certificate
 
 (LOCAL (DEFUN CERT (A B C X)
          (+ (MONOID-CF-0 A B C X)
             (CONE-CF-0 A B C X)
             (* (IDEAL-CF-0 A B C X) (PROB-0 A B C X)))))

;; Contradictory results on the sign of the certificate
 
 (LOCAL (DEFTHMD CERT-KEY
          (IMPLIES (AND (RATIONALP A)
                        (RATIONALP B)
                        (RATIONALP C) (RATIONALP X))
                   (= (CERT A B C X) 0))
          :hints
          (("Goal" :in-theory
                   (enable SQUARE
                           CERT
                           PROB-0
                           PROB-1
                           IDEAL-CF-0 CONE-CF-0 MONOID-CF-0)))))
 
 (LOCAL (DEFTHM CERT-CONTRA-M-0
          (IMPLIES (AND (NOT (GOAL A B C X))
                        (RATIONALP A)
                        (RATIONALP B)
                        (RATIONALP C) (RATIONALP X))
                   (> (MONOID-CF-0 A B C X) 0))
          :hints
          (("Goal" :in-theory
                   (enable SQUARE
                           CERT
                           PROB-0
                           PROB-1
                           IDEAL-CF-0 CONE-CF-0 MONOID-CF-0)))
          :rule-classes (:linear)))
 
 (LOCAL (DEFTHM CERT-CONTRA-C-0
          (IMPLIES (AND (NOT (GOAL A B C X))
                        (RATIONALP A)
                        (RATIONALP B)
                        (RATIONALP C) (RATIONALP X))
                   (>= (CONE-CF-0 A B C X) 0))
          :rule-classes (:linear)))
 
 (LOCAL (DEFTHM CERT-CONTRA-I-0
          (IMPLIES (AND (NOT (GOAL A B C X))
                        (RATIONALP A)
                        (RATIONALP B)
                        (RATIONALP C) (RATIONALP X))
                   (= (* (IDEAL-CF-0 A B C X)
                         (PROB-0 A B C X))
                      0))
          :hints
          (("Goal" :in-theory
                   (enable SQUARE
                           CERT
                           PROB-0
                           PROB-1
                           IDEAL-CF-0 CONE-CF-0 MONOID-CF-0)))
          :rule-classes (:linear)))
 
 (LOCAL (DEFTHM CERT-CONTRA
          (IMPLIES (AND (NOT (GOAL A B C X))
                        (RATIONALP A)
                        (RATIONALP B)
                        (RATIONALP C) (RATIONALP X))
                   (NEQ (CERT A B C X) 0))
          :rule-classes nil))

;; Main lemma
 
 (LOCAL (DEFTHM MAIN
          (IMPLIES (AND (RATIONALP A)
                        (RATIONALP B)
                        (RATIONALP C) (RATIONALP X))
                   (GOAL A B C X))
          :hints
          (("Goal" :in-theory
                   (disable GOAL)
                   :use (CERT-KEY CERT-CONTRA)))
          :rule-classes nil))

;; Final theorem
 
 (DEFTHM FINAL
   (IMPLIES (AND (RATIONALP A)
                 (RATIONALP B)
                 (RATIONALP C)
                 (RATIONALP X) (= (+ (* A X X) (* B X) C) 0))
            (>= (- (* B B) (* 4 A C)) 0))
   :hints
   (("Goal" :in-theory
            (enable GOAL PROB-0 PROB-1) :use (MAIN)))
   :rule-classes nil))
\end{minted}

\section{Mathematical Background}\label{sec:psatz}

The general setting for nonlinear real arithmetic is the theory of real closed fields (RCF). A real closed field is a field elementarily equivalent to $\mathbb{R}$ w.r.t. the language of ordered rings, i.e., the first-order language of polynomial equations and inequalities over $\mathbb{Q}[\vec{x}]$.
RCF is complete, decidable and admits effective elimination of quantifiers~\cite{tarski-rcf-qe,passmore-phd-thesis}.

Though decidable, RCF is fundamentally infeasible. For example, Davenport-Heintz have isolated a family of $n$-variable RCF formulas of length $O(n)$ whose only quantifier-free equivalents must contain polynomials of degree $2^{2^{\Omega(n)}}$ and of length $2^{2^{\Omega(n)}}$~\cite{davenport-heintz}. Tarski was the first to give an RCF quantifier elimination algorithm~\cite{tarski-rcf-qe} but its non-elementary complexity makes it impractical for real-world use. Collins's CAD~\cite{collins-cad} achieves an asymptotic best-case of doubly-exponential complexity and is the foundation of many best performing proof procedures available in computer algebra systems and SMT solvers~\cite{passmore-phd-thesis}. Nevertheless, CAD relies on complex algebro-geometric computations and to date no one has succeeded in extracting foundationally checkable proof objects from CAD.

For restricted fragments of RCF, we can do better. The purely existential fragment is known to only have singly exponential worst-case complexity~\cite{canny}, and convex optimization techniques can efficiently handle many specialized but practically useful classes of problems~\cite{parrilo-sdp}. It is in this context that our work takes place: we are working only over the purely universal (dually, purely existential) fragment, and our proof construction uses convex optimization to search over a space of possible foundational proofs.

\subsection{The Krivine-Stengle Positivstellensatz}

The core of our proof construction relies on the Krivine-Stengle Positivstellensatz. Like its complex algebro-geometric sibling the Nullstellensatz, the Positivstellensatz guarantees the existence of algebraic proof certificates witnessing unsatisfiability. While the Nullstellensatz deals only with equations and ideals and their relationship with satisfiability over $\mathbb{C}$, the Positivstellensatz is more intricate as it must also take into account ordering relations given $\mathbb{R}$'s status as an ordered field. 

\begin{theorem}[Krivine-Stengle Positivstellensatz]
\[
\left( \bigwedge_i^{k_0} p_i = 0\right) \wedge 
\left( \bigwedge_i^{k_1} q_i \geq 0\right) \wedge 
\left( \bigwedge_i^{k_2} r_i \neq 0\right)
\ \ \text{ s.t. } \ \
p_i, q_i, r_i \in \mathbb{Q}[\vec{x}]
\]
is unsatisfiable over $\mathbb{R}$ iff
\[
\exists \mathrm{P} \in Ideal(p_1, \mathellipsis, p_{k_0})
\]
\[\exists \mathrm{Q} \in Cone(q_1, \mathellipsis, q_{k_1})\]
\[
\exists \mathrm{R} \in Monoid(r_1, \mathellipsis, r_{k_2})\\
\]
s.t.
\[
\mathrm{P} + \mathrm{Q} + \mathrm{R}^2 = 0
\]
where
\[
Ideal(a_1, \mathellipsis, a_m) = \left\{\sum_{i=1}^m a_i b_i \ | \ b_i \in \mathbb{Q}[\vec{x}]\right\}
\]
\[
Cone(a_1, \mathellipsis, a_m) = \left\{r + \sum_{i=1}^m t_i u_i \ | \ r, t_i \in \sum(\mathbb{Q}[\vec{x}])^2, u_i \in Monoid(a_1, \mathellipsis, a_m)\right\}
\]
\[
Monoid(a_1, \mathellipsis, a_m) = \left\{\prod_{i=1}^m (a_i)^j \ | \ j \in \mathbb{N}\right\}
\]
\[
\sum(\mathbb{Q}[\vec{x}])^2 = \left\{ \sum_{i=1}^v (p_i)^2 \ | \ p_i \in \mathbb{Q}[\vec{x}] \ \wedge \ v \in \mathbb{N} \right\}.
\]
\end{theorem}

The sum $\mathrm{P} + \mathrm{Q} + \mathrm{R}^2$ is the certificate of unsatisfiability. Like we reasoned in the introduction, it is easy to see why unsatisfiability follows: appealing to the fact that ideals generalize nullity, cones generalize non-negativity, and multiplicative monoids generalize non-nullity, our constraints imply that $\mathrm{P}=0$, $\mathrm{Q}\geq0$ and $\mathrm{R}^2>0$, and thus that $\mathrm{P} + \mathrm{Q} + \mathrm{R}^2 > 0$. But by polynomial arithmetic alone $\mathrm{P} + \mathrm{Q} + \mathrm{R}^2$ reduces to $0$. Thus our constraint system implies $0>0$ and must be unsatisfiable. 
The miracle of the theorem is that these certificates always exist. The next question is: how to find them?

\subsection{Sums of Squares Decompositions and Semidefinite Programming}

From the guise of logic, the Positivstellensatz gives us both a proof system and a completeness theorem. The original proofs establishing the Positivstellensatz, however, were non-constructive, giving no hint as to how one can effectively find the promised proofs. 

A major advance occurred in 2000, with Parrilo's use of semidefinite programming (SDP) relaxations to efficiently search over convex spaces of certificate coefficients~\cite{parrilo-phd-thesis,parrilo-sdp}. 

From our perspective, Parillo's key theorem is the following (Theorem 5.1 of~\cite{parrilo-phd-thesis}):

\begin{theorem*}[SDP for Positivstellensatz Search]
Consider a system of polynomial equalities and inequalities. Then, the search for bounded degree Positivstellensatz refutations can be done using semidefinite programming. If the degree bound is chosen to be large enough, then the SDPs will be feasible, and the certificates obtained from its solution.
\end{theorem*}

The critical fact is that these searches are over a convex space, and thus can take advantage of efficient optimization methods. How to make the space convex? For a given certificate bound, we consider which monomials could possibly appear in the certificate, and introduce fresh  variables for them. Then, the problem polynomials can be expressed as a quadratic form in the fresh variables, and a linear constraint system (\emph{modulo} a PSD constraint on the matrix of the quadratic form) can be extracted by comparing coefficients. But optimizing linear constraints modulo a PSD matrix is a convex optimization problem: this is precisely the domain of semidefinite programming. 

\section{Examples, Caveats and Limitations}

While we believe our present approach is promising and useful in many ways, especially for relatively small but algebraically nontrivial inequalities arising in verification practice, it is not a panacea.
First, the worst-case degree bounds on certificates are in general hyper-exponential in dimension, and we experience this in practice: the more variables there are, the harder things tend to get. Second, the space of possible certificates grows rapidly as degree bounds are expanded. And third, efficient SDP solvers use numerical methods based on floating point, and it is not always easy to recover exact rational coefficients from SDP solutions. 
Harrison's {\tt REAL\_SOS} tactic~\cite{harrison-sos} in HOL-Light addresses many of these challenges, and we refer the reader to his work for more details. Subsequent theoretical analyses have shown that some of these issues are insurmountable with the present approach~\cite{monniaux-psatz-degenerate}.

Nevertheless, we are encouraged by the present state of the method. For example, the following are all problems which can be solved by Imandra, translated into ACL2, and checked successfully by ACL2 in (at most) seconds:

\begin{minted}{lisp}
 (IMPLIES (= (+ (* X X) (* Y Y) (* Z Z)) 1)
          (<= (* (+ X Y Z) (+ X Y Z)) 3))

 (IMPLIES (= (+ (* W W) (* X X) (* Y Y) (* Z Z)) 1)
          (<= (* (+ W X Y Z) (+ W X Y Z)) 4))

 (IMPLIES (AND (<= 0 X) (<= 0 Y) (= (* X Y) 1))
          (<= (+ X Y) (+ (* X X) (* Y Y))))

 (IMPLIES (AND (>= X 1) (>= Y 1))
          (>= (* X Y) (- (+ X Y) 1)))

 (IMPLIES (AND (<= 0 X) (<= 0 Y))
               (<= (* X Y (EXPT (+ X Y) 2))
                   (EXPT (+ (* X X) (* Y Y)) 2)))

 (IMPLIES (AND (<= 0 A) (<= 0 B) (<= 0 C) 
               (<= (* C (EXPT (+ (* 2 A) B) 3)) (* 27 X)))
           (<= (* C A A B) X))
\end{minted}

There are some problems which, e.g., Harrison's {\tt REAL\_SOS} can handle, but we cannot. We are not sure why, but we conjecture this may have to do with numerical differences in the execution of the SDP solver, as SDP floating point results can be platform dependent~\cite{harrison-sos}. These include:

\begin{minted}{lisp}
 (IMPLIES (AND (= (+ (* A X X X) (+ B X X) (+ C X) D) 0)
               (= (+ (* A Y Y Y) (+ B Y Y) (+ C Y) D) 0)
               (< (+ (- (* 18 A B C D) (* 4 B B D))
                  (- (* B B C C) (* 4 A C C C))
                  (- 0 (* 27 A A D D)))
                0))
          (= X Y))
\end{minted}
and
\begin{minted}{lisp}
 (IMPLIES (AND (= (- X2 U3) 0)
               (= (* (- (- X1 U1) U3) (* X2 U2)) 0)
               (= (- (* X4 X1) (* X3 U3)) 0)
               (= (- (* X4 (- U2 U1)) (* (- X3 U1) U3)) 0))
          (= (+ (- (- (* X1 X1) (* 2 X1 X3)) (* 2 X4 X2)) (* X2 X2)) 0))
\end{minted}
Encouragingly, in all such failing cases, we fail even to construct a proof in Imandra, rather than finding a certificate but failing in extracting a valid ACL2 proof. In all of our current examples, if we find a certificate, we successfully construct an ACL2 version which ACL2 checks quickly.

\section{Related Work}

Harrison's HOL-Light {\tt REAL\_SOS} tactic~\cite{harrison-sos} is the moral foundation of this work.
For the case of Positivstellensatz proofs, we have in many ways simply adapted his ideas to the setting of Imandra and ACL2, including his OCaml interface to the {\tt csdp}~\cite{csdp} SDP solver and techniques for rational certificate recovery. Harrison's work is based on Parillo's key insight of reducing Positivstellensatz searches to a sequence of SOS decompositions~\cite{parrilo-sdp}, which in turn builds on the Powers-Wörmann algorithm for reducing SOS decompositions to a sequence of convex SDP searches~\cite{powers-wormann,passmore-sos-explained}. 

\section{Conclusion and Future Work}
We have presented an integration of Imandra and ACL2 for constructing ACL2 proofs of nonlinear inequalities.
The approach is built around the Positivstellensatz and uses convex optimization to search for foundational proofs of unsatisfiability.
This work is in many ways an Imandra and ACL2 adaptation of the pioneering work of Harrison and his {\tt REAL\_SOS} tactic in HOL-Light, and further of Parrilo's work on reducing Positivstellensatz searches to semidefinite programming.
We are next focusing on integrating Imandra's real algebraic counterexample search and region decomposition methods into the procedure~\cite{passmore-imandra,passmore-demoura-infinitesimal}, and further handling problems with more general boolean structure.
We also aim to develop an ACL2 client (available in, e.g., Emacs) which makes it easy to send problems to an Imandra service in the cloud and to then incorporate the delivered proofs into local developments.

\bibliographystyle{eptcs}
\bibliography{arith}
\end{document}